\documentclass[final,5p,times,twocolumn, 
]{elsarticle}
\usepackage{mathrsfs}
\usepackage{amssymb}
\usepackage{amsmath}
\usepackage{graphicx}
\usepackage{amsfonts,amsbsy}
\usepackage{nicefrac}
\usepackage{xcolor}
\usepackage{soul}
\usepackage[breaklinks,colorlinks,urlcolor=blue,citecolor=citcolor,linkcolor=lcolor]{hyperref}
\usepackage{array}
\usepackage{multirow}
\usepackage{siunitx}

\definecolor{lcolor}{rgb}{0.5,0,0}
\definecolor{citcolor}{rgb}{0,0.3,0.0}
\definecolor{ao(english)}{rgb}{0.0, 0.5, 0.0}
\definecolor{RoyalBlue}{HTML}{0071BC} 
\usepackage{comment}

\newcommand{\mrm}{\mathrm}
\newcommand{\Q}{Q_s}
\newcommand{\dA}{d_{\mrm{A}}}
\newcommand{\dR}{d_{\mrm{R}}}
\newcommand{\CA}{C_{\mrm{A}}}
\newcommand{\NC}{N_\mathrm{c}}
\newcommand{\CF}{C_{\mrm{F}}}

\newcommand{\fig}{Fig.~}

\newcommand{\eq}{Eq.~}

\newcommand{\app}{}

\newcommand{\qperp}{q_\perp}

\newcommand{\qhat}{\hat q}
\newcommand{\qhatf}{\qhat_{\mathrm{f}}}
\newcommand{\qhatff}{\qhat_{\mathrm{ff}}}

\newcommand{\vb}{\vec}
\renewcommand{\vec}[1]{\mathrm{\mathbf{#1}}}
\newcommand{\dd}[2][]{\mathrm d^{#1}{#2}\,} 
\newcommand{\dv}[2][]{\frac{\dd{#1}}{\dd{#2}}}
\newcommand{\pdv}[2][]{\frac{\partial{#1}}{\partial{#2}}}

\newcommand{\Ejet}{E_{\mathrm{jet}}}

\newcommand{\taubmss}{\tau_{\mathrm{BMSS}}}

\newcommand{\Conetwo}{\mathcal {C}^{1\leftrightarrow 2}}
\newcommand{\Ctwotwo}{\mathcal{ C}^{2\leftrightarrow 2}}
\newcommand{\Cexp}{\mathcal C^{\mathrm{exp}}}
\newcommand{\Teps}{T_{\varepsilon}}

\DeclareSIUnit\c{c}

\journal{Physics Letters B}

\begin{document}
\begin{frontmatter}
\title{Jet momentum broadening during initial stages in heavy-ion collisions}

\author[a]{K.~Boguslavski} 
\affiliation[a]{
    organization={Institute for Theoretical Physics, TU Wien},
    addressline={Wiedner Hauptstrasse 8-10},
    city={Vienna},
    postcode={1040},
    country={Austria}
    }

\author[b]{A.~Kurkela} 
\affiliation[b]{organization={Faculty of Science and Technology, University of Stavanger, 4036 Stavanger, Norway}}

\author[c,d]{T.~Lappi} 
\affiliation[c]{organization={Department of Physics, University of Jyväskylä, P.O.~Box 35, 40014 University of Jyväskylä, Finland}}
\affiliation[d]{organization={Helsinki Institute of Physics, P.O.~Box 64, 00014 University of Helsinki, Finland}}

\author[a]{F.~Lindenbauer}
\corref{Corresponding author}
\ead{florian.lindenbauer@tuwien.ac.at}

\author[c,d,e]{J.~Peuron} 
\affiliation[e]{organization={Department of Physics, Lund University,  Sölvegatan  14A, Lund, SE-223 62, Sweden}}
\begin{abstract}

We study the jet quenching parameter $\hat q$ in the initial pre-equilibrium stages of heavy-ion collisions using the QCD kinetic theory description of the anisotropic quark-gluon plasma.
This allows us to smoothly close the gap in the literature between the early glasma stage of the collision and the onset of hydrodynamics. We find that 
the pre-hydrodynamic evolution of $\qhat$ during the bottom-up kinetic scenario shows little sensitivity to the initial conditions, jet energies and models of the transverse momentum cutoff.
We also observe that, similarly to the glasma case, the jet quenching parameter is 
enhanced along the beam axis as compared to the transverse direction during most of the kinetic evolution. 
\end{abstract}


\end{frontmatter}

\section{Introduction}
Jets are important probes of the quark-gluon plasma generated in heavy-ion collisions \cite{PHENIX:2001hpc, PHENIX:2004vcz, STAR:2002svs, STAR:2005gfr, ATLAS:2010isq, ALICE:2010yje, CMS:2011iwn}. They originate from a highly energetic quark or gluon created in the initial hard collision \cite{Qin:2015srf, Apolinario:2022vzg}
, and result in hadrons with large momentum measured in the detector.
The jet quenching parameter $\qhat$ describes transverse momentum broadening of this leading 
parton along its trajectory
and encodes medium effects to the jet evolution and to energy loss \cite{Baier:1998yf, Majumder:2013re, Blaizot:2015lma,Cao:2021rpv, Apolinario:2022vzg}.

Due to their early creation, jets probe all phases of the quark-gluon plasma evolution, including the earliest stages. These involve 
the glasma, a phase shortly after the collision that is driven by highly occupied classical gluonic fields, followed by the kinetic theory stage, in which the system is described as an interacting gas of gluon and quark quasiparticles, before eventually a hydrodynamical description becomes applicable, and local equilibrium is reached (see e.g.~\cite{Schlichting:2019abc, Berges:2020fwq} for reviews). 
For a thermal medium, the jet quenching parameter $\qhat$ can be calculated for weak \cite{Arnold:2008vd, Caron-Huot:2008zna} and strong coupling \cite{Liu:2006ug, Zhang:2012jd} using perturbative techniques and lattice Quantum Chromodynamics (QCD) \cite{Kumar:2020wvb}, and recent progress has been made to describe jet-medium interactions in thermal equilibrium \cite{Ghiglieri:2015ala, Moore:2021jwe, Ghiglieri:2022gyv, Mehtar-Tani:2022zwf, Song:2022wil, Grishmanovskii:2022tpb}.  
Also extractions of $\qhat$ from experimental data commonly use a thermal or hydrodynamic background for the medium evolution \cite{JET:2013cls, JETSCAPE:2021ehl, Xie:2022ght}. 

Jet quenching is considered to be a dominant effect for several experimental observables, such as the nuclear modification factor $R_{AA}$ or $v_2$ at high $p_T$. Addressing these requires models such as those of Refs.~\cite{Schenke:2009gb, Zapp:2012ak, Casalderrey-Solana:2014bpa, Cao:2017hhk,Andres:2019eus, Zigic:2019sth, Andres:2022bql, Huss:2020whe,JETSCAPE:2021ehl, Zhao:2021vmu, JETSCAPE:2022jer, Xie:2022ght, JETSCAPE:2023hqn} that integrate over the whole collision process. Currently, such models typically only incorporate the effects on jets of a thermal, hydrodynamical medium after a finite starting time. However, the precise treatment of jet quenching during the initial pre-equilibrium stages has not been fully understood.
For instance, while in Ref.~{\cite{Andres:2019eus}} $\qhat$ is required to be suppressed before ${\SI{0.6}{\femto\meter/c}}$ for the simultaneous description of $R_{AA}$ and $v_2$, in Ref.~{\cite{JETSCAPE:2023hqn}} this is not the case. Additionally, in the glasma phase
large values 
and direction dependent momentum broadening have been reported \cite{Ipp:2020mjc, Ipp:2020nfu, Carrington:2021dvw, Carrington:2022bnv, Avramescu:2023qvv}.
This discrepancy shows the need for a 
consistent description of the evolution of $\qhat$, which requires the knowledge of its evolution throughout the whole  pre-equilibrium stage.
This can be achieved in the weak coupling bottom-up thermalization scenario~\cite{Mueller:1999pi, Baier:2000sb, Berges:2013eia, Berges:2013fga, Kurkela:2015qoa}
using QCD effective kinetic theory (EKT)
\cite{Arnold:2002zm}, which is commonly extrapolated to moderate couplings to study the approach to hydrodynamics{\cite{Kurkela:2015qoa, Kurkela:2018wud, Kurkela:2018xxd, Du:2020zqg}}.
Although $\qhat$ and jet-medium interactions have been discussed for anisotropic \cite{Romatschke:2004au, Romatschke:2006bb, Dumitru:2007rp, Baier:2008js, Hauksson:2021okc, Hauksson:2023tze}, inhomogeneous and flowing systems \cite{Sadofyev:2021ohn,Barata:2022utc, Andres:2022ndd, Barata:2022krd},
so far, no computation during the  bottom-up equilibration dynamics in heavy-ion collisions \cite{Baier:2000sb} exists. A short description of the stages of the bottom-up thermalization scenario is found in
\ref{app:bottom-up}. 

\begin{figure*}
\centerline{
\raisebox{0.55cm}{\includegraphics[width=0.4\linewidth]{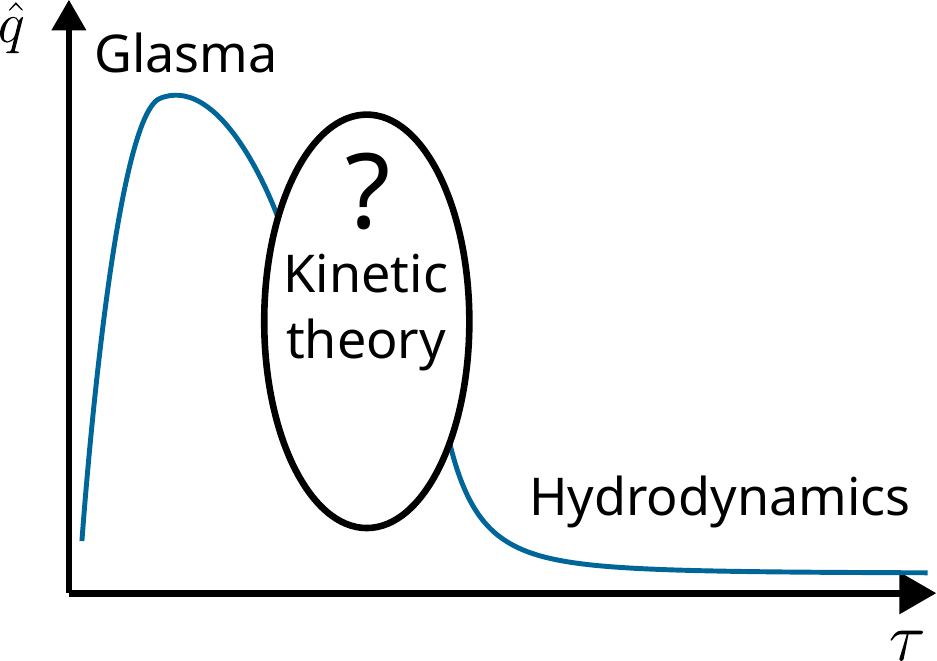}}
  \includegraphics[width=0.45\linewidth]{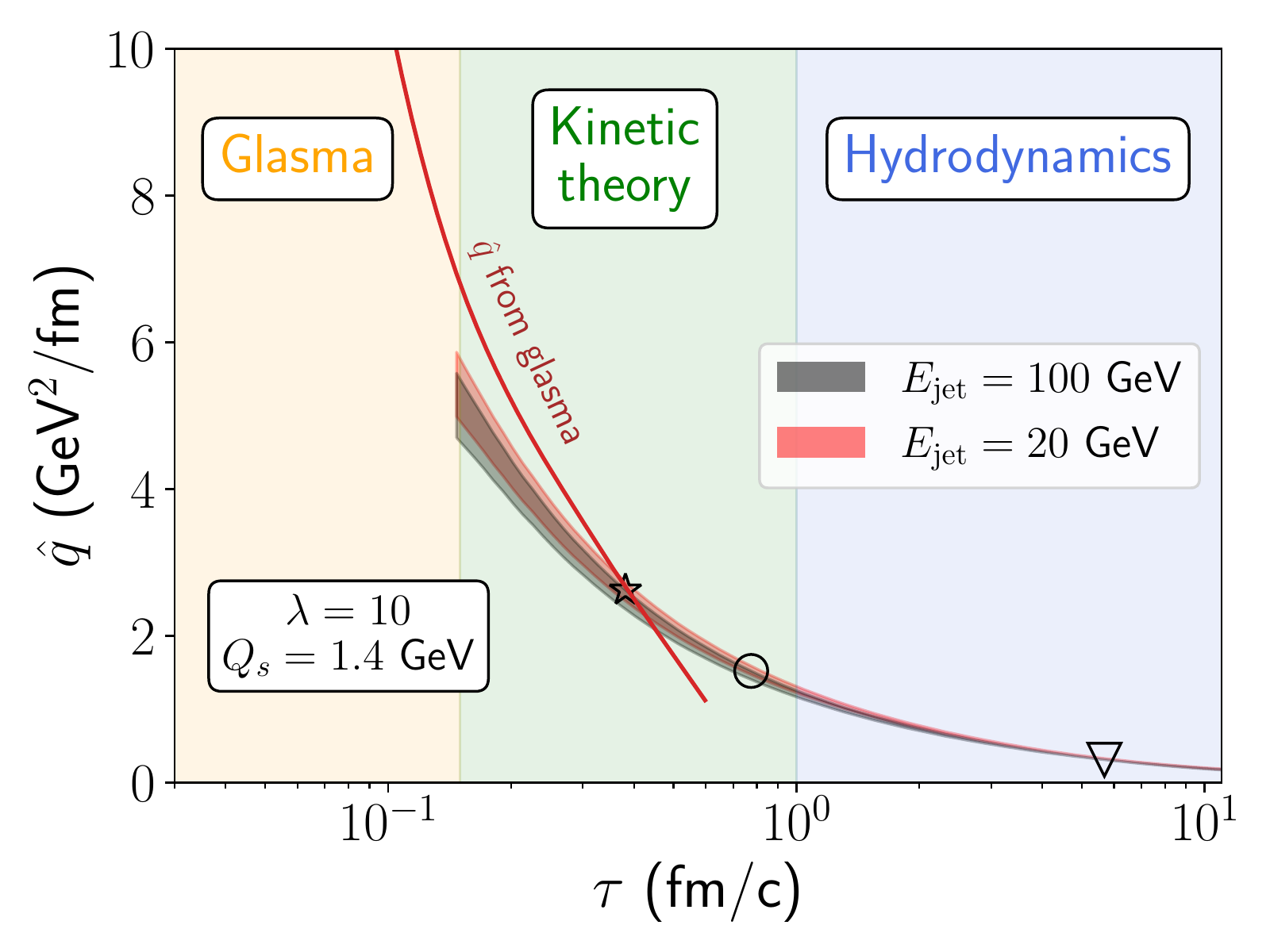}
  }
\caption{\label{fig:qhat_appetizer_combinedv2}
(Left): Schematic evolution of $\qhat$ during the different stages of the plasma evolution. (Right): 
The values of $\qhat$ relevant for a quark jet for $\lambda=10$, $\Q=\SI{1.4}{\giga\electronvolt}$ and jet energies $\Ejet=\SI{20}{\giga\electronvolt}$ (red band) and $\SI{100}{\giga\electronvolt}$ (black band).
Each band includes simulation results for different initial conditions with parameters $\xi = 4, 10$ and the time-dependent transverse momentum cutoff models of Eq.~\eqref{eq:cutoff_combined}. 
The markers indicate different stages of the bottom-up thermalization as explained in the text (see also
\ref{app:bottom-up}).
The glasma calculation of Ref.~\protect{\cite{Ipp:2020nfu}} is shown as a red line for comparison.
}
\end{figure*}

In this letter, we tackle this problem by computing
$\qhat$ between the glasma and the onset of hydrodynamics using kinetic theory, as sketched in the left panel of \fig\ref{fig:qhat_appetizer_combinedv2}. 
We find that the large values of $\qhat$ reported from the glasma consistently connect to the hydrodynamical values at later times. As shown for 
simulations relevant for jet energies $E_{\mathrm{jet}}=20-\SI{100}{\giga\electronvolt}$ in 
the right panel of 
\fig\ref{fig:qhat_appetizer_combinedv2}, this provides remarkable agreement with the previously reported glasma values of $\qhat$ that were argued to have a sizable impact on total jet energy loss \cite{Carrington:2021dvw, Ipp:2020nfu, Avramescu:2023qvv}.
We will discuss our approach and our steps towards this result in the remainder of this work.

\section{Kinetic theory}
To study the plasma evolution during bottom-up thermalization, we perform numerical simulations within QCD effective kinetic theory (EKT) \cite{Arnold:2002zm} for a gluonic plasma, since gluons are the dominant degrees of freedom at early times \cite{Kurkela:2018xxd, Kurkela:2018oqw}. Using the same setup as in \cite{Kurkela:2015qoa}, we describe the plasma in terms of 
quasiparticle distribution functions $f_{\vb p}\equiv f(\tau,\vb p)$, whose evolution in proper time $\tau$ is given by the Boltzmann equation
\begin{align}
-\pdv[f_{\vb p}]{\tau}=\Conetwo[f_{\vb p}]+ \Ctwotwo [f_{\vb p}]+\mathcal \Cexp[f_{\vb p}].\label{eq:boltzmann_equation}
\end{align}
Here $\Ctwotwo$ encodes collisions resulting in elastic scatterings, $\Conetwo$ describes effective splitting terms relevant for gluon radiation, and $\Cexp = - \frac{p_z}{\tau} \pdv[f_{\vb p}]{p_z}$ accounts for the longitudinal expansion of the plasma in the beam direction \cite{Mueller:1999pi, Kurkela:2015qoa}. Our only free parameter 
is the 't Hooft coupling $\lambda=\NC g^2$, where $\NC=3$ is the rank of the Yang-Mills gauge group. The effective kinetic theory description \cite{Arnold:2002zm} is leading-order accurate for isotropic distributions but neglects the effect of plasma instabilities that may be present in anisotropic systems \cite{Mrowczynski:1993qm, Kurkela:2011ti, Kurkela:2011ub}. We use an isotropic screening prescription that is typically employed in EKT implementations \cite{AbraaoYork:2014hbk, Kurkela:2015qoa, Kurkela:2018xxd, Kurkela:2018oqw, Du:2020dvp}.

\section{Jet quenching parameter}
To study jet momentum broadening, we consider elastic scatterings of the leading jet parton with momentum $\vb p$ off a plasma constituent with momentum $\vb k$.
Their momenta are $\vb p'$ and $\vb k'$ after the collision with a momentum transfer $\vb q = (q_x,\vb q_\perp)=\vb p' - \vb p$. To account for different directions, we define the jet quenching parameter matrix
\begin{align}
    \qhat^{ij}=\dv[\langle q^iq^j\rangle]{L},
\end{align}
where $L$ is the length along the trajectory of the jet.
We consider the jet moving in the $x$ direction where $z$ is the beam axis and thus distinguish transverse momentum broadening via $\qhat^{yy}$ and $\qhat^{zz}$. The usual (isotropic) jet quenching parameter is the sum $\qhat=\qhat^{yy}+\qhat^{zz}$. In kinetic theory, we calculate $\qhat^{ij}$ 
for a gluonic background using the perturbative expression \cite{Boguslavski:2023waw} 
\begin{align}
&\hat q^{ij}({\tau}) = \frac{1}{4\dR}\lim_{|\vb p|\to\infty}\int_{\substack{\vb k\vb k'\vb p'\\q_\perp < \Lambda_\perp}} \!\!\!q_\perp^i q_\perp^j (2\pi)^4\delta^3(\vb p+\vb k-\vb p'-\vb k') \nonumber\\
&\qquad\times\delta(|\vb p|+|\vb k|-|\vb p'|-|\vb k'|)\frac{\left|\mathcal M_{ag}^{ag}\right|^2}{|\vb p|} f_{\vb k}\left(1+ f_{\vb k'}\right),
\label{eq:qhat_general}
\end{align}
with $\int_{\vb k}=\int\frac{\dd[3]{\vb k}}{(2\pi)^3 2|\vb k|}$ and where $\dR$ is the dimension of the representation of the jet. In our numerical calculations, we consider quark jets with $\dR = \NC = 3$. The values of $\qhat$ for gluons can be obtained  via Casimir scaling $\qhat^{\mathrm{gluon}}/\CA=\qhat^{\mathrm{quark}}/\CF$ with $\CA=3$, $\CF=4/3$ for QCD. 
The matrix element $\left|\mathcal M^{ag}_{ag}\right|^2$ describes the elastic scattering of the jet parton $a$ off a gluon in the plasma and is calculated in leading-order perturbative QCD.
For any internal soft line we use the hard thermal loop (HTL) matrix element \cite{Arnold:2002zm} with the Debye mass calculated within the simulations. 
Similarly as in the EKT time evolution, in the matrix element for $\qhat$ we employ an isotropic screening approximation, neglecting the effect of plasma instabilities, which were shown not to play a dominant role during the thermalization dynamics {\cite{Berges:2013eia, Berges:2013fga}}.
Although all results in this paper are obtained using this matrix element, we have also checked numerically that an often employed screening approximation \cite{AbraaoYork:2014hbk, Kurkela:2015qoa, Kurkela:2018oqw, Kurkela:2018xxd, Du:2020dvp} does not lead to significant changes (see \app \ref{app:screening_approximation}).

The parameter $\qhat$ is defined only up to a cutoff scale $\Lambda_\perp$, whose natural value has been argued to depend on the jet energy $\Ejet$ and the effective medium temperature \cite{Caron-Huot:2008zna,He:2015pra,JETSCAPE:2021ehl, Mehtar-Tani:2022zwf}.
For instance in thermal equilibrium, the jet quenching parameter can be calculated for large cutoffs $\Lambda_\perp\gg T$ as \cite{Arnold:2008vd, Caron-Huot:2008zna}
\begin{align} 
\qhat_{\mathrm{therm}}(\Lambda_\perp\gg T) = \lambda^2 T^3\, \frac{C_R \zeta(3)}{\NC\pi^3} \left(\ln\frac{\Lambda_\perp}{\sqrt{\lambda}\,T}+\text{ const}\right). \label{eq:qhat_thermal}
\end{align}
For $\Lambda_\perp\sim T$---as well as for systems out of equilibrium---the dependence is more complicated but can be obtained from Eq.~\eqref{eq:qhat_general}. Our code implementation can be found in \cite{kurkela_2023_10409474}.

We want to compare our non-equilibrium simulation results to a ``corresponding'' thermal equilibrium. 
To do this, we use the Landau matching condition and
define an energy density matched temperature as
\begin{align}
    \Teps(\tau)=\left(\frac{30\,\varepsilon(\tau)}{\pi^2 \nu_g}\right)^{1/4}
    \label{eq:def_Teps}
\end{align}
 obtained from the energy density of our nonequilibrium system $\varepsilon = \nu_g\int\frac{\dd[3]{\vb p}}{(2\pi)^3} |\vb p| f_{\vb p}$. Here $\nu_g = 2(\NC^2-1)$ counts the number of gluonic spin and color degrees of freedom. 

Due to its leading proportionality to $\lambda^2 \Teps^3$, we show our results for $\qhat^{ij}$ scaled by this factor. Moreover,
we use the scale 
\begin{equation}
\taubmss = 
\alpha_\mathrm{s}^{-13/5}\Q^{-1}
=
\left(\frac{\lambda}{4\pi\NC}\right)^{-13/5}\Q^{-1},
\end{equation}
to rescale our time variable since it parametrically captures the thermalization time during the kinetic bottom-up scenario \cite{Baier:2000sb, Kurkela:2015qoa}.


\begin{figure}
\centering
\includegraphics[width=\linewidth]{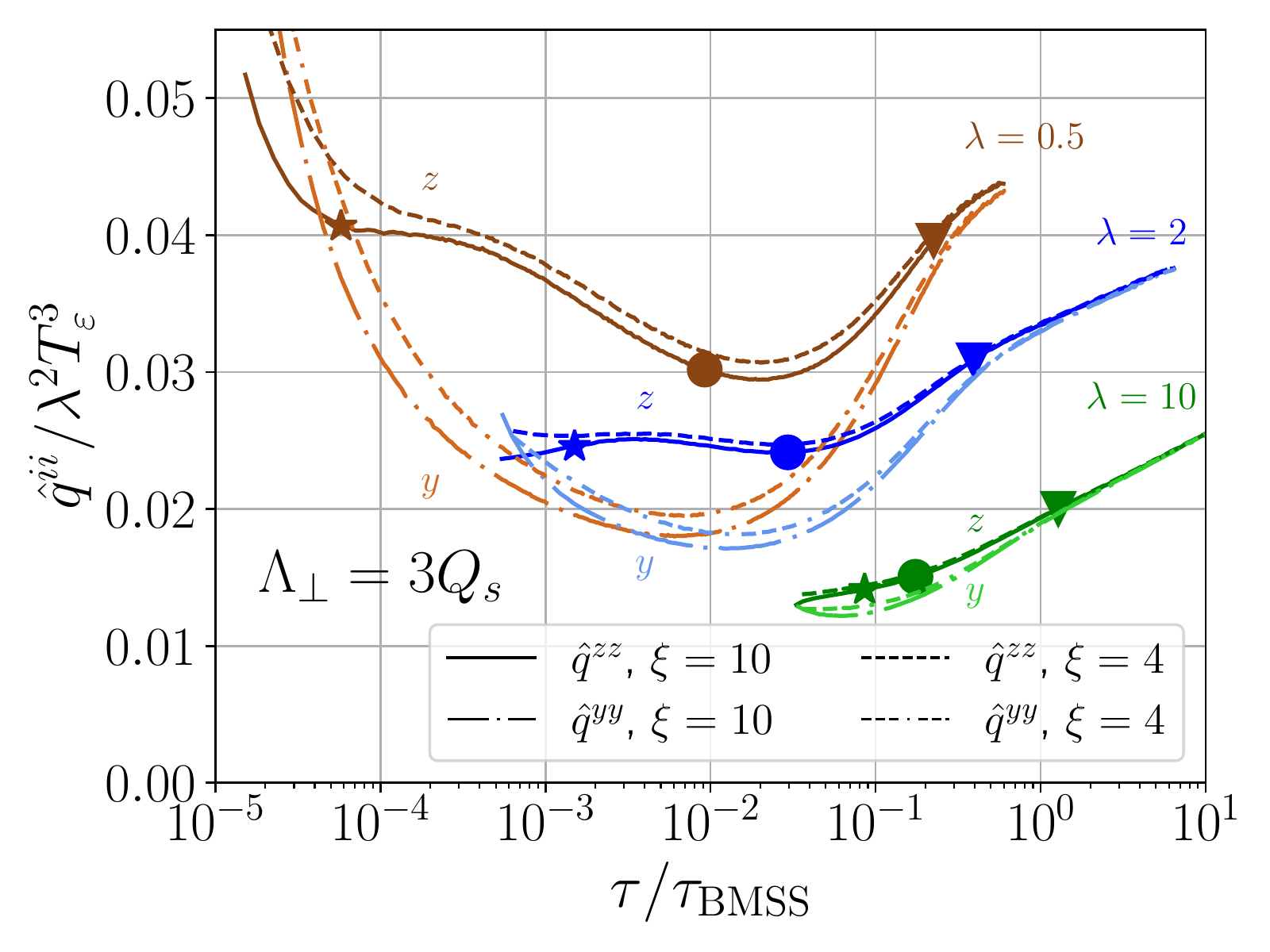}
\caption{\label{fig:qhat_couplings_aniso}
Jet quenching parameter in $y$ and $z$ directions rescaled by $\lambda^2 \Teps^3$ during bottom-up thermalization for momentum cutoff $\Lambda_\perp=3\Q$ for different couplings and initial conditions (solid: $\xi=10$, dashed: $\xi=4$). 
}
\end{figure}

\section{Initial conditions and time markers}
We initialize the simulations by choosing the same initial distribution at time $\Q\tau=1$ 
as in \cite{Kurkela:2015qoa}%
\begin{align}
	\begin{split}
		&f({\tau=1/\Q,}p_\perp,p_z)=\frac{2}{\lambda}A \frac{\langle p_T\rangle}{\sqrt{p_\perp^2+(\xi p_z)^2}}\\
		&\qquad\times\exp\left(-\frac{2}{3\langle p_T\rangle^2}\left(p_\perp^2+(\xi p_z)^2\right)\right).
	\end{split}\label{eq:bottom_up_initcond}
\end{align}
We use two sets of initial parameters with different
momentum anisotropies $\xi$ and $\langle p_T \rangle  = 1.8 \Q$,
\begin{align}
\label{eq:initcond}
\xi=10, \quad A = 5.24171;
\qquad
\xi = 4, \quad A = 2.05335,
\end{align}
where $A$ is chosen as in {\cite{Kurkela:2015qoa}} with $\tau\epsilon=\langle p_T\rangle \dd{N}/\dd[2]{\vb x}\dd{y}$ fixed via the total gluon multiplicity per unit rapidity $\dd{N}/\dd[2]{\vb x}\dd{y}$ in the glasma from {\cite{Lappi:2011ju}}.
Once initialized, a weakly coupled system follows the bottom-up thermalization scenario that consists of three stages \cite{Baier:2000sb, Kurkela:2015qoa}: an anisotropic highly occupied plasma, a radiative stage of approximately constant anisotropy, and a final thermalization stage via an inverse energy cascade (summarized in \app \ref{app:bottom-up}). 
To guide the eye, we roughly separate these stages by special time markers. The star marker is placed where the occupancy is $f \sim 1/\lambda$, the circle marker lies at the minimum occupancy, and the triangle marker indicates where the pressure ratio is $P_T/P_L=2$, which marks a time close to equilibrium.

\begin{figure}
  \centering
    \includegraphics[width=\linewidth]{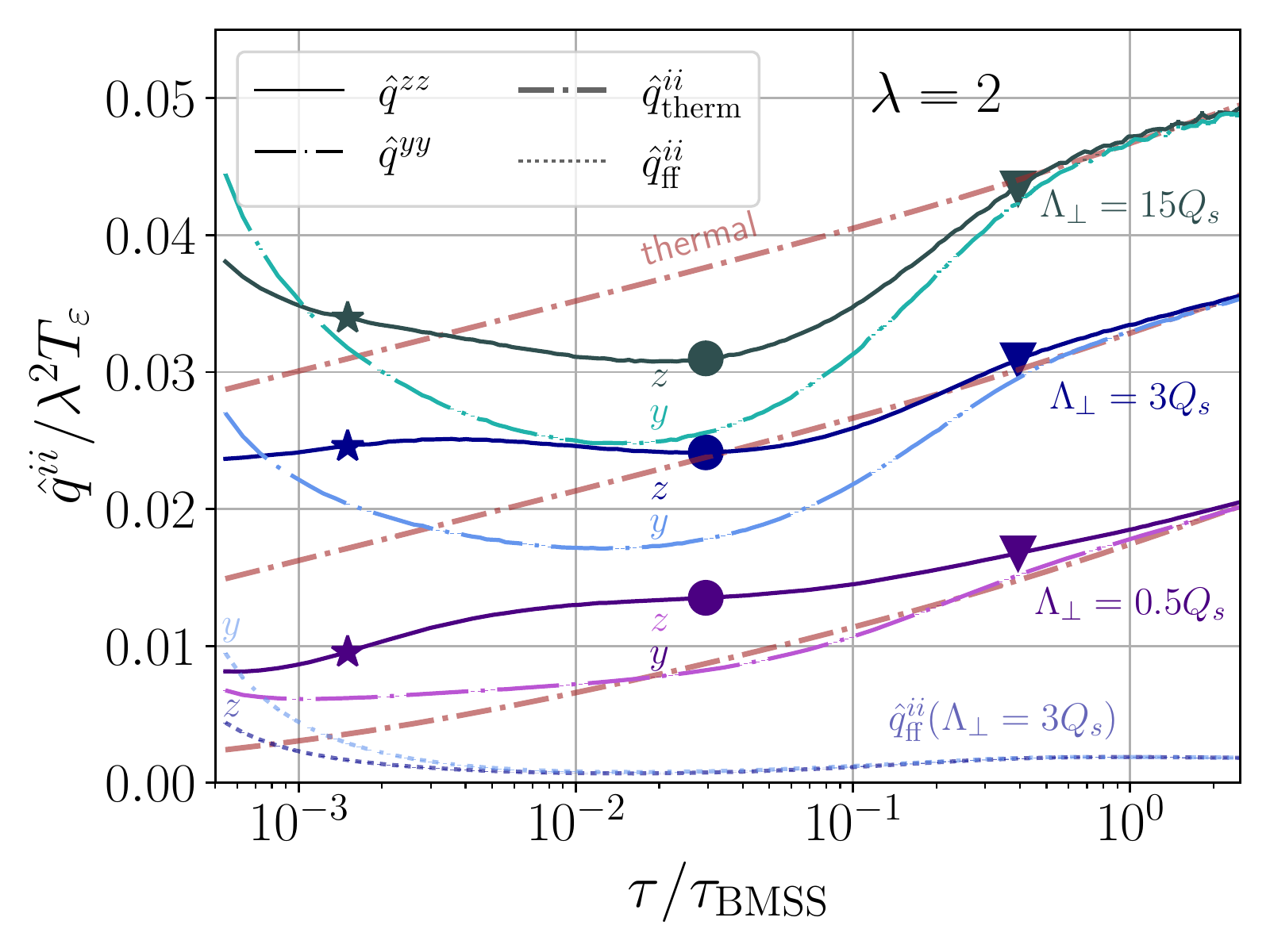}
    \caption{
    \label{fig:qhat_approach_to_thermal}
    Evolution of $\qhat^{zz}$ and $\qhat^{yy}$ for a quark jet during bottom-up thermalization for $\lambda=2$ and different cutoffs $\Lambda_\perp$. We have added the Bose enhanced contributions $\qhatff^{ii}$ for $\Lambda_\perp = 3\Q$ as dotted curves of the same color, visible at the bottom of the figure.
    For comparison, thermal curves for the same $\Teps(\tau)$ are shown as brown dash-dotted lines. 
}
\end{figure}

\section{Simulation results}
We start by discussing the resulting momentum broadening coefficient at fixed transverse momentum cutoffs $\Lambda_\perp$ and later generalize this to more realistic models of evolving momentum cutoffs. 
While mixed components $\qhat^{yz}=\qhat^{zy} = 0$ vanish in our simulations in accordance with symmetry arguments, 
we study here the evolution of $\qhat^{zz}$ and $\qhat^{yy}$. 
In \fig\ref{fig:qhat_couplings_aniso} we show them at the cutoff $\Lambda_\perp = 3 \Q$ for different couplings and initial anistropy parameters $\xi = 10$ (solid lines) and $\xi = 4$ (dashed lines). We find little sensitivity to the variation of initial conditions, less than $15\%$ for the considered parameters. Moreover, we observe qualitatively similar behavior for different couplings. 

To further study the  evolution of $\qhat^{ii}$, we show their values for different cutoffs $\Lambda_\perp$ in \fig \ref{fig:qhat_approach_to_thermal} for the anisotropy $\xi = 10$ and coupling $\lambda = 2$, and find qualitatively similar effects also for others couplings.
The  estimates for an energy-density matched (see Eq.~\eqref{eq:def_Teps}) thermal system $\qhat^{ii}_{\mathrm{therm}}=\frac{1}{2}\,\qhat_{\mathrm{therm}}$
are also shown as dash-dotted lines and are obtained by evaluating Eq.~\eqref{eq:qhat_general} with a thermal distribution. We also show separately the contribution from the Bose-enhanced $\qhatff$ term as dotted lines (defined as the part of Eq.~\eqref{eq:qhat_general} that is proportional to $f^2$).
We observe that in general, the order of magnitude of  $\qhat^{ii}$ follows the energy-density matched thermal values. In the earliest stage of bottom-up thermalization, characterized by overoccupation and extreme anisotropy and marked by the star symbols,  the values  of $\qhat$ are above the energy-density matched thermal ones.  At the next stage, marked by the circles, the values for large cutoffs $\Lambda_\perp$ undershoot the thermal ones, while those for a small cutoff overshoot them. This behavior results from low occupancies of the plasma constituents, as we have verified by studying a scaled thermal distribution. Finally, approaching thermal equilibrium (the triangle markers), the values of $\qhat^{ii}$ also approach the thermal expectation.

For almost the entire evolution we find that momentum broadening in the beam direction is larger than transverse to it, $\qhat^{zz}>\qhat^{yy}$. 
This seems to be typical for anisotropic systems with occupancies up to order unity,
as has been found for transport coefficients in the context of kinetic theory \cite{Romatschke:2006bb, Dumitru:2007rp, Boguslavski:2023fdm}. 
In our formulation this ordering is a result of the anisotropic under-occupied distribution and thus, does not stem from the matrix element for which an isotropic HTL screening prescription is used.
It leads to a sizable difference in the total momentum broadening in different directions. 
Moreover, a low momentum cutoff can be associated to momentum broadening of the plasma constituents themselves. Thus the larger broadening in the $z$ direction for smaller $\Lambda_\perp$ is consistent with the isotropization dynamics in the bottom-up scenario.

Interestingly, we find that for large cutoffs this ordering is reversed at early times before the star marker, leading to $\qhat^{zz}<\qhat^{yy}$. This mainly stems from the Bose-enhancement of the over-occupied plasma phase at the beginning of the evolution. To make this more quantitative, we have plotted the Bose-enhancement part $\qhatff$ of $\qhat$ in \fig\ref{fig:qhat_approach_to_thermal}, separately for the $y$ and $z$ directions. Note that $\qhatff$ is finite in the limit $\Lambda_\perp \to \infty$, and the value at $\Lambda_\perp=3\Q$ that is plotted is already close to that limit. While for the non-Bose enhanced term $\qhatf$, the anisotropy $p_z \ll p_\perp$ leads to $\qhatf^{zz}>\qhatf^{yy}$, for the Bose-enhanced term the effect is the opposite, $\qhatff^{zz}<\qhatff^{yy}$.
This, coupled with the larger occupation numbers in the earliest stage, leads to the observed initially reversed anisotropy of $\qhat$.

Remarkably, jet quenching studies in the glasma \cite{Ipp:2020mjc, Ipp:2020nfu, Avramescu:2023qvv} have revealed a similar ordering $\qhat^{zz}>\qhat^{yy}$ as we find for most of the evolution of $\qhat$ in our kinetic simulations, although for a different reason. There the enhancement of $\qhat^{zz}$ seems to stem primarily from a slight asymmetry between the chromo-magnetic and -electric fields in the underlying classical-statistical description of the glasma.


\section{Results for realistic cutoff dependence}
Until now we have studied $\qhat$ using a fixed cutoff $\Lambda_\perp$. This is 
unphysical since during the expansion all characteristic energy scales of the plasma decrease. To account for this, 
we choose 
cutoff models that depend on the jet energy $\Ejet$ and
effective
plasma temperature
$\Teps$:
\begin{subequations}\label{eq:cutoff_combined}
\begin{align}
    \Lambda_\perp^{\mathrm{LPM}}(\Ejet,\Teps)&=\zeta^{\mathrm{LPM}}g\times(\Ejet \Teps^3)^{1/4}\label{eq:cutoff_LPM}\\
    \Lambda_\perp^{\mathrm{kin}}(\Ejet,\Teps)&=\zeta^{\mathrm{kin}}g\times(\Ejet \Teps)^{1/2}.\label{eq:cutoff_kinematic}
\end{align}
\end{subequations}
The first cutoff model $\Lambda_\perp^{\mathrm{LPM}}$ is a rough estimate of the accumulated transverse momentum during the formation time of a gluon emission during the LPM regime, where quantum mechanical interference leads to a suppression of the emission rate. It can be obtained from estimates of the relevant formation time $t^{\mathrm{form}} \sim \Ejet/\qperp^2$, $\qhat \sim g^4 T^3$ and $\qperp^2 \sim \qhat t^{\mathrm{form}}$ \cite{Arnold:2008zu, Arnold:2008vd, Caron-Huot:2008zna, Kurkela:2014tla}.
Variants of the kinematic
cutoff model $\Lambda_\perp^{\mathrm{kin}}$ have been widely used in the literature \cite{Qin:2009gw, JET:2013cls, Xu:2014ica, He:2015pra, Cao:2021rpv, JETSCAPE:2021ehl,JETSCAPE:2022jer, Mehtar-Tani:2022zwf} and take into account that the plasma particles the jet scatters off have momentum $k\sim T$.
We emphasize that we regard $\qhat$ as a medium property relevant for a jet with an appropriate fixed energy $\Ejet$ and do not study the actual evolution of a jet. 

In principle, the cutoff is process dependent. 
However, no substantial differences are expected as we will show below, since the dependence of $\qhat$ on the cutoff is only logarithmic for large $\Lambda_\perp$. 
\begin{align}
    \qhat^{ii}(\Lambda_\perp\gg \Teps)\simeq a_i\ln\frac{\Lambda_\perp}{\Q}+b_i. \label{eq:qhat_form_large_cutoff}
\end{align}
In practice, we fit the coefficients $a_i$ and $b_i$ to the large cutoff behavior of the numerically obtained values for $\qhat$. 
The numerical values of $a_i/\Q^3$ and $b_i/\Q^3$ as functions of $\Q\tau$ for $\lambda=0.5, 1, 2, 5$ and $10$ for the initial conditions in \eq\eqref{eq:initcond} are provided in \cite{lindenbauer_2023_10419537}. 

To make comparisons with the glasma, we set our $\Q$ such that we reproduce the energy density of the glasma in Ref.~\cite{Ipp:2020nfu} at our initial time $\Q\tau=1$ %
\footnote{Note that the value of $\Q^{\mathrm{glasma}}$ used in glasma simulations might not correspond to the value of $\Q$ we use due to different definitions and conventions. 
For the glasma simulation in Ref.~\cite{Ipp:2020nfu}, $\Q^{\mathrm{glasma}}=\SI{2}{\giga\electronvolt}$ and $m/(g^2\mu)=0.1$ with $50$ color sheets have been used, where $g^2\mu$ and $\Q^{\mathrm{glasma}}$ are related as in Ref.~\cite{Lappi:2007ku}.}.
This yields a value of $\Q=\SI{1.4}{\giga\electronvolt}$. This is the same value as obtained in Ref.~\cite{Keegan:2016cpi} that is needed 
for the EKT setup to be consistent with the later hydrodynamic evolution. This shows the consistency of both approaches. We then obtain the values of the parameters $\zeta$ in \eq\eqref{eq:cutoff_combined}
by matching them 
at the triangle marker close to thermal equilibrium.  Concretely, we match the values at this close-to-equilibium point where
$\Teps=0.21\Q=\SI{295}{\mega\electronvolt}$ and realistic coupling $\lambda=10$ to the median value for $\qhat_{\mathrm{therm}}$ in the LBT parametrization of the JETSCAPE collaboration \cite{JETSCAPE:2021ehl}, in order to be close to a conventional numerical estimate for a thermal distribution. For $\Ejet=\SI{20}{\giga\electronvolt}$ we obtain $\zeta^{\mathrm{LPM}}=0.70$ and $\zeta^{\mathrm{kin}}=0.16$, whereas for $\Ejet=\SI{100}{\giga\electronvolt}$ we obtain $\zeta^{\mathrm{LPM}}=1.14$ and $\zeta^{\mathrm{kin}}=0.40$.
In \app{\ref{app:different_matchings}} we show that matching to other models does not significantly change our results.

\begin{figure}
  \centering
  \includegraphics[width=\linewidth]{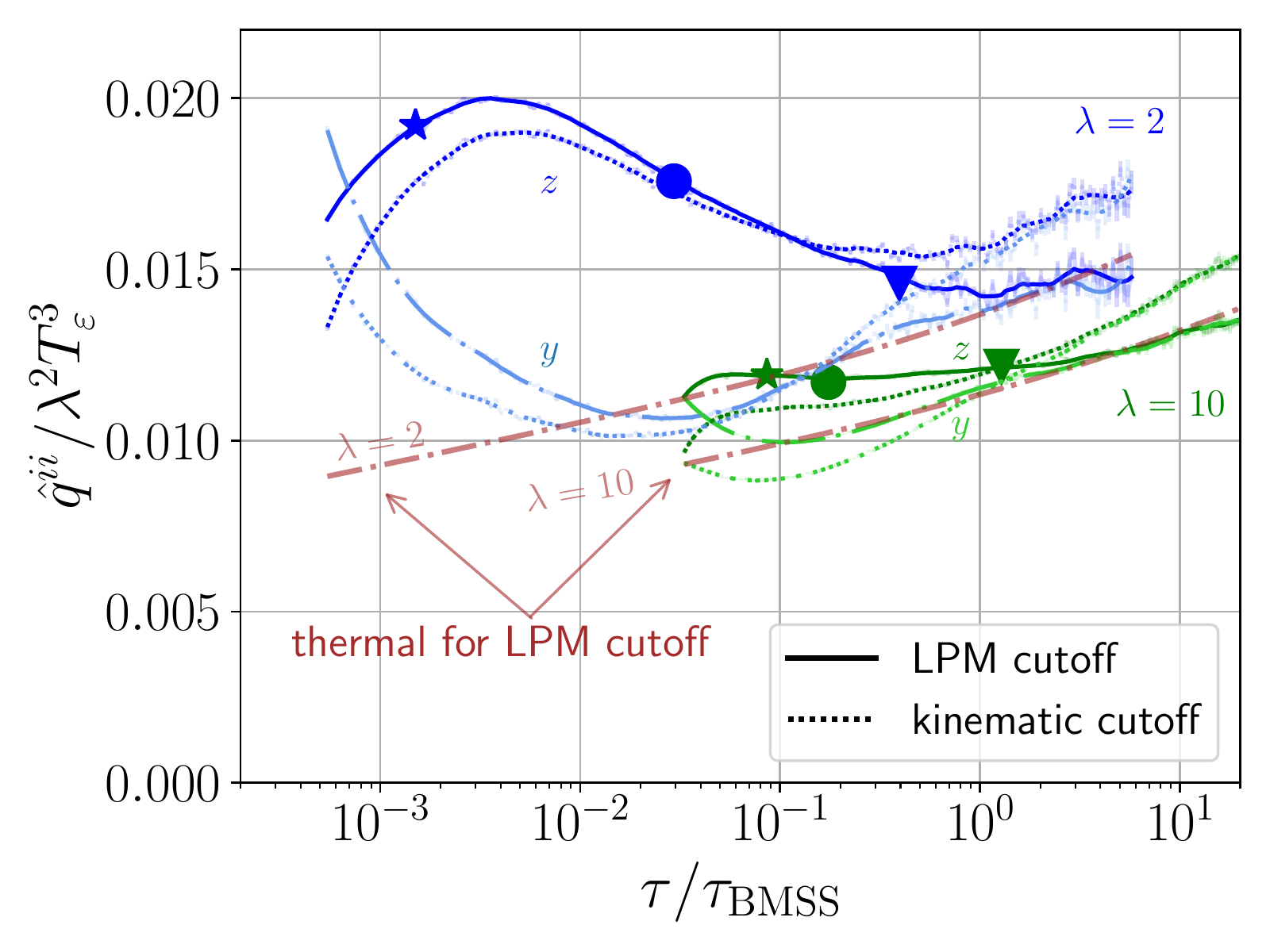}
  \caption{\label{fig:qhat_realisticv5a}
  Evolution of $\qhat^{zz}$ and $\qhat^{yy}$ for the cutoff models in \eqref{eq:cutoff_LPM} (solid) and \eqref{eq:cutoff_kinematic} (dashed) with jet energy $\Ejet=\SI{100}{\giga\electronvolt}$, $\Q=\SI{1.4}{\giga\electronvolt}$
  and $\Teps$ extracted from the plasma simulation for $\xi=10$. 
  The curves were smoothed using a Savitzky-Golay filter, while the original curves with estimated error bars are shown transparently beneath. 
  Thermal curves for the LPM cutoff model are included for comparison.
  }
\end{figure}

We show $\qhat^{zz}$ and $\qhat^{yy}$ in \fig\ref{fig:qhat_realisticv5a} for the anisotropy parameter $\xi=10$, jet energy $\Ejet = \SI{100}{\giga\electronvolt}$, different values of the coupling $\lambda$, and both cutoff parametrizations of \eq \eqref{eq:cutoff_combined}.
We observe a similar evolution of $\qhat^{ii}$ as for fixed cutoffs, with $\qhat^{zz}>\qhat^{yy}$ for most of the non-equilibrium evolution except for a short transient phase of reversed ordering at the beginning for $\lambda = 2$. We also find that longitudinal momentum broadening is more efficient than expected from a thermal system, as indicated by the dashed thermal line for comparison.
Both cutoff models lead to similar results, with the LPM cutoff yielding systematically higher values than the kinematic cutoff model during the pre-hydrodynamic evolution. For $\lambda = 10$ they differ by less than $20\%$, while the variation of initial anisotropies has a much smaller impact, as we have seen in \fig\ref{fig:qhat_couplings_aniso}. 

This relatively mild sensitivity to the initial parameters and cutoff models
enables us to predict the value of $\qhat$ throughout the pre-equilibrium stages, extrapolating backwards from a fit to a phenomenological extraction by the JETSCAPE collaboration at the triangle time marker close to equilibrium. This is the procedure that leads to the numerical values shown at the beginning of this work, 
in the right panel of 
\fig\ref{fig:qhat_appetizer_combinedv2} for 
jet energies $\Ejet = {\SI{20}{\giga\electronvolt}}$ and $\SI{100}{\giga\electronvolt}$, where each band contains simulations with both cutoff models and both initial conditions. The overlapping bands signal little sensitivity to the jet energy.
At $\tau \sim \SI{0.2}{\femto\meter}/c$ we find $\qhat \approx 4 - \SI{5}{\giga\electronvolt^2/\femto\meter}$. As visible in the figure, this is comparable to the values during the glasma regime at this time and thus provides a connection to the earliest phase of the plasma evolution. 
The ratio $\qhat^{yy}/\qhat^{zz}$ is included in \app \ref{app:aniso_qhat} and shows a sizable suppression of up to 20\% between the glasma and the hydrodynamic regimes.

\section{Conclusions}
In this letter, we have studied the jet quenching parameter $\qhat$ at initial stages in heavy-ion collisions. Often characterizing jet energy loss, 
$\qhat$ at early times is a major uncertainty in theory predictions for experimental observables.
In contrast to phenomenological studies that find $\qhat$ initially suppressed {\cite{Andres:2019eus}} or neglect jet energy loss during initial stages \cite{Schenke:2009gb, He:2015pra, Zapp:2012ak, Casalderrey-Solana:2014bpa}, we obtain significant values of $\qhat$ during its pre-equilibrium evolution using QCD kinetic theory simulations, consistently connecting the glasma results with hydrodynamics.

Our $\qhat$ results follow roughly Landau-matched thermal estimates to a first approximation while exhibiting a sizable anisotropy $\qhat^{zz}>\qhat^{yy}$ for most of the pre-equilibrium evolution. The latter shows the importance of going beyond equilibrium approximations since these systematically underestimate $\qhat$ and particularly $\qhat^{zz}$.

We then moved from a fixed transverse momentum cutoff to one
In our analysis, we studied the impact of different couplings, initial conditions, and transverse momentum cutoffs.
In particular, we employed a varying cutoff that depends on the evolving energy density of the medium, through commonly used LPM and kinematic models, and
obtained a relatively parameter- and model-independent evolution of $\qhat$.
For that, we matched 
its numerical value
at the late, close-to-equilibium region to phenomenological extractions from heavy-ion collision data. Extrapolating backwards, we were able to obtain a result for $\qhat$ in physical units that we extended to 
early times
where it matched surprisingly well with earlier estimates from classical field simulations of the glasma stage. This matching gives us confidence that we are starting to obtain a realistic, consistent estimate of the jet quenching parameter $\qhat$ throughout the pre-equilibrium stages.

Our results of $\qhat$ during the kinetic regime could be used to extend current frameworks that employ a hydrodynamic medium evolution to extract $\qhat$ from experimental data \cite{JET:2013cls, JETSCAPE:2021ehl}. 
Although based on scattering processes with on-shell partons, our extracted values for $\qhat$ can also enter jet evolution models in order to include medium effects during the initial large virtuality phase \cite{JETSCAPE:2023hqn}.
Since the anisotropy $\qhat^{zz} > \qhat^{yy}$ remains during most of the pre-hydrodynamic evolution including the glasma and kinetic stages, it may have observable consequences, such as a sizable jet hadron polarization as suggested in \cite{Hauksson:2023tze}.


\section*{Acknowledgments}

We would like to thank S.\ Hauksson, A.\ Ipp, J.G.\ Milhano, D.I.\ M\"uller, and M.\ Strickland for valuable discussions.
We are particularly grateful to D.I.\ M\"uller for providing his data on the glasma results.
KB and FL would like to thank the Austrian Science Fund (FWF) for support under project P 34455, and FL is additionally supported by the Doctoral Program W1252-N27 Particles and Interactions.
TL and JP are supported by the Academy of Finland, the Centre of Excellence in Quark Matter (project 346324)
and project 321840  and  by the European Research Council under project ERC-2018-ADG-835105 YoctoLHC.
This work was also supported under the European Union’s
Horizon 2020 research and innovation  by the STRONG-2020 project (grant agreement No. 824093). The content of this article does not reflect the official opinion of the European Union and responsibility for the information and views expressed therein lies entirely with the authors.
The computational results presented have been achieved in part using the Vienna Scientific Cluster (VSC), project 71444.

\appendix

\begin{figure}[t]
    \centering
    \includegraphics[width=\linewidth]{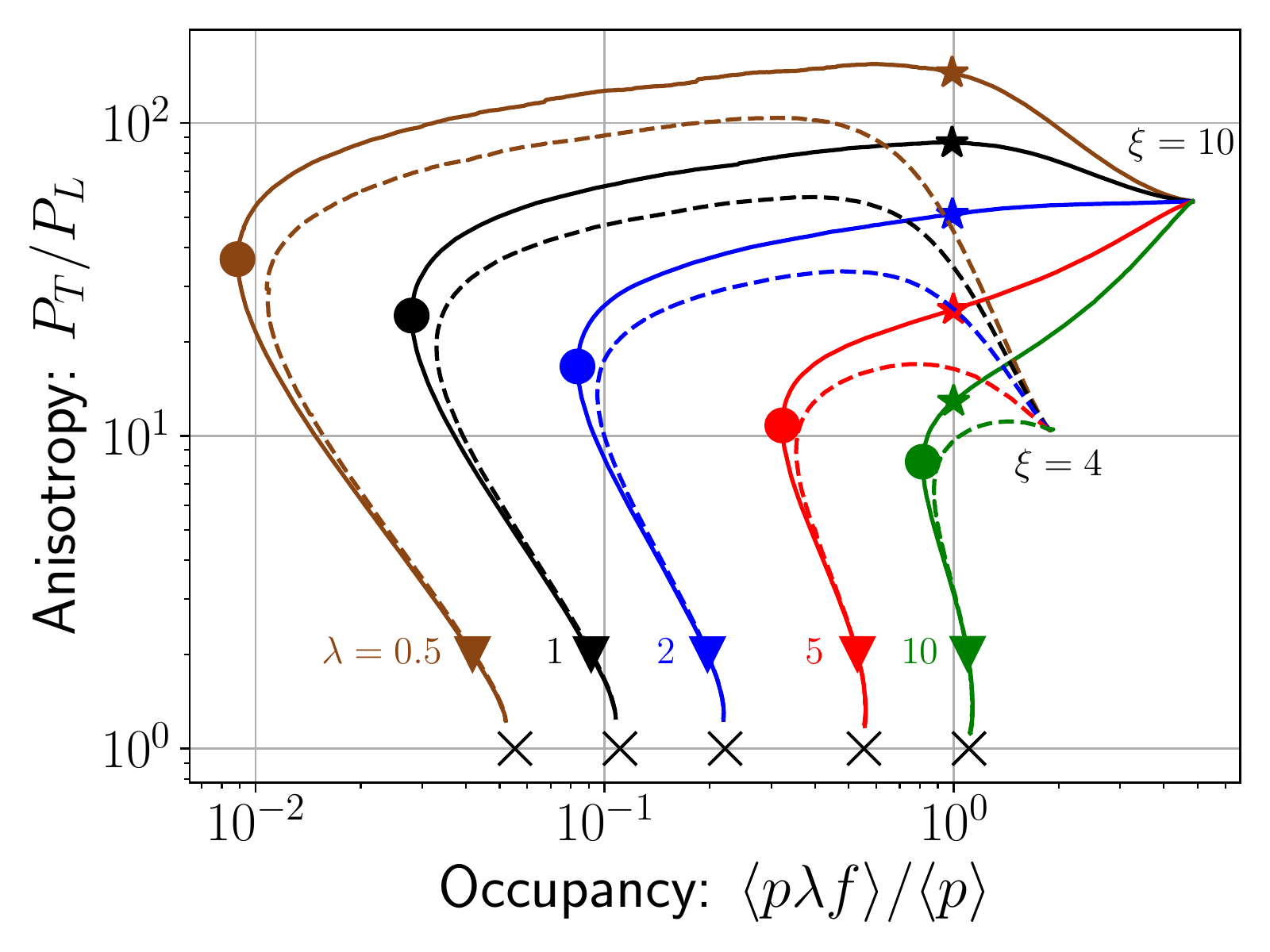}
    \caption{\label{fig:overview_curves}
    EKT simulations for different initial conditions and different couplings. The star time marker is placed at occupancy $\langle p\lambda f\rangle/\langle p\rangle = 1$, the circle marker at minimum occupancy, the triangle marker where $P_T/P_L=2$ and the cross indicates thermal equilibrium.
    }
\end{figure}

\section{Bottom-up thermalization}
\label{app:bottom-up}

The pre-thermal quark-gluon plasma generated in heavy-ion collisions follows the bottom-up thermalization scenario \cite{Mueller:1999pi, Baier:2000sb, Berges:2013eia, Kurkela:2015qoa}, which consists of several stages:
The first stage is dominated by a large number of hard gluons, and the anisotropy increases due to the longitudinal expansion along the beam axis. When the occupancy of these hard modes reaches unity, we enter the second stage, where a significant amount of soft gluons is produced through branching while the momentum anisotropy remains roughly constant. These soft gluons form a thermal bath, but a significant amount of the  total energy is still carried by the remaining small number of hard gluons. In the third stage, these hard gluons lose energy through multiple hard branchings, until they join the thermal bath and the system equilibrates.

Numerical simulations \cite{Kurkela:2015qoa} support this picture. In \fig\ref{fig:overview_curves} we show the   time evolution of the system in the anisotropy-occupancy plane, where we also introduce three markers to guide the eye that roughly correspond to the boundaries between the three stages described above. The star marker is placed where the occupancy is $\langle pf \rangle / \langle p \rangle = 1/\lambda$, the circle is placed at the minimum occupancy, and the triangle marker indicates where the pressure ratio is $P_T/P_L=2$, and marks the time close to equilibrium.

The longitudinal $P_L$ and transverse pressure $P_T$ are obtained from the corresponding components of the energy-momentum tensor (with $K^0=|\vb k|$)
\begin{align}
&T^{\mu \nu} = 2\nu_g \int \frac{\dd[3]{\vb k}}{(2\pi)^3 2|\vb k|} K^\mu K^\nu  f(\vb k) \\
&P_T = \frac{T^{xx}+T^{yy}}{2}\,, \qquad P_L = T^{zz}.
\end{align}
The occupancy of the hard modes can be assessed via
\begin{align}
    \frac{\langle p\lambda f\rangle}{\langle p\rangle}=\lambda\frac{\int {\dd[3]{\vb p}}|\vb p| f^2(\vb p)}{\int \dd[3]{\vb p}|\vb p| f(\vb p)}.
\end{align}


\begin{figure}[t]
    \includegraphics[width=\linewidth]{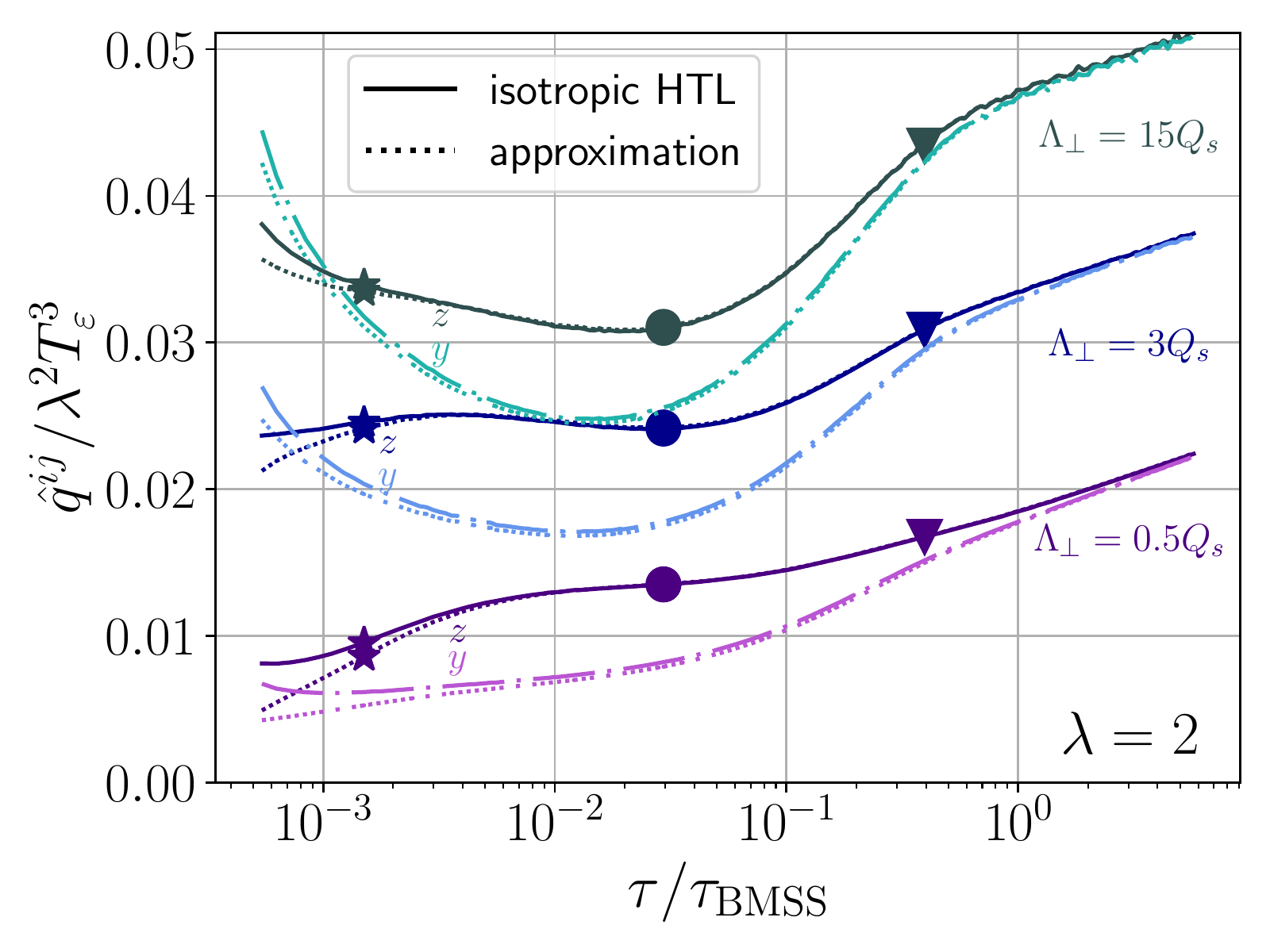}
    \caption{
    \label{fig:qhat_comparison_HTL}
    Comparison between $\qhat^{ii}$ for different matrix elements.
    Solid lines: full isotropic HTL matrix element;
     dashed lines: approximated matrix element using the simple screening form \eqref{eq:screened_isotropic_matrix_element}.
     }
\end{figure}

\section{Screening approximation to the matrix element\label{app:screening_approximation}}

For the evaluation of $\qhat$, we use matrix elements with soft gluon exchanges regulated by the isotropic HTL self-energy, as described in Ref.~{\cite{Arnold:2002zm}}. We start with the matrix element for gluon scattering in vacuum
\begin{align}
    {\frac{\left|\mathcal M^{gg}_{gg}\right|^2}{4\dA\lambda^2}=9+\frac{(s-t)^2}{u^2}+\frac{(t-u)^2}{s^2} +\frac{(u-s)^2}{t^2},}
\end{align}
and take the eikonal limit $|\vb p|\to\infty$, in which only the last term contributes. Medium screening effects are incorporated via the replacement (see {\cite{Arnold:2002zm}})
\begin{align}
    {\frac{(s-u)^2}{t^2}\to \to\left|G_R(P-P')_{\mu\nu}(P+P')^\mu (K+K')^\nu\right|^2,}
\end{align}
where for $G_R$ we use the isotropic form of the hard-thermal loop propagator in Coulomb gauge. The resulting expression is lengthy but can be straightforwardly obtained and is listed in Ref.~\cite{Boguslavski:2023waw}.

An often-used approximation 
is a simple screening form \cite{AbraaoYork:2014hbk, Kurkela:2015qoa, Kurkela:2018xxd, Kurkela:2018oqw, Du:2020dvp}, including a screening parameter $\xi_\perp = e^{1/3}/2$ for transverse momentum broadening
\begin{align}
\begin{split}
    &\lim_{|\vb p|\to\infty}\frac{\left|\mathcal M^{gg}_{gg}\right|^2}{4\dA\lambda^2\vb p^2}\\
    &\quad =4\frac{\left(2|\vb k|-\omega -\sqrt{(2|\vb k|-\omega)^2-\vb q^2}\cos\phi_{kq}\right)^2}{(\vb q^2+\xi_\perp^2m_D^2)^2}.
\end{split} \label{eq:screened_isotropic_matrix_element}
\end{align}
In all figures of the paper, we use the full isotropic HTL resummed matrix element. 
Additionally, we check the validity of the simple screening approximation \eqref{eq:screened_isotropic_matrix_element} here and show a comparison between $\qhat^{ii}$ in \fig \ref{fig:qhat_comparison_HTL} for both matrix elements. One observes only minor differences, mostly visible at small cutoffs $\Lambda_\perp$ and at early times. This makes the simple screening approximation in \eqref{eq:screened_isotropic_matrix_element} applicable to the jet quenching parameter during bottom-up thermalization.

\section{Different cutoff matchings\label{app:different_matchings}}
\begin{figure}
    \centering
    \includegraphics[width=\linewidth]{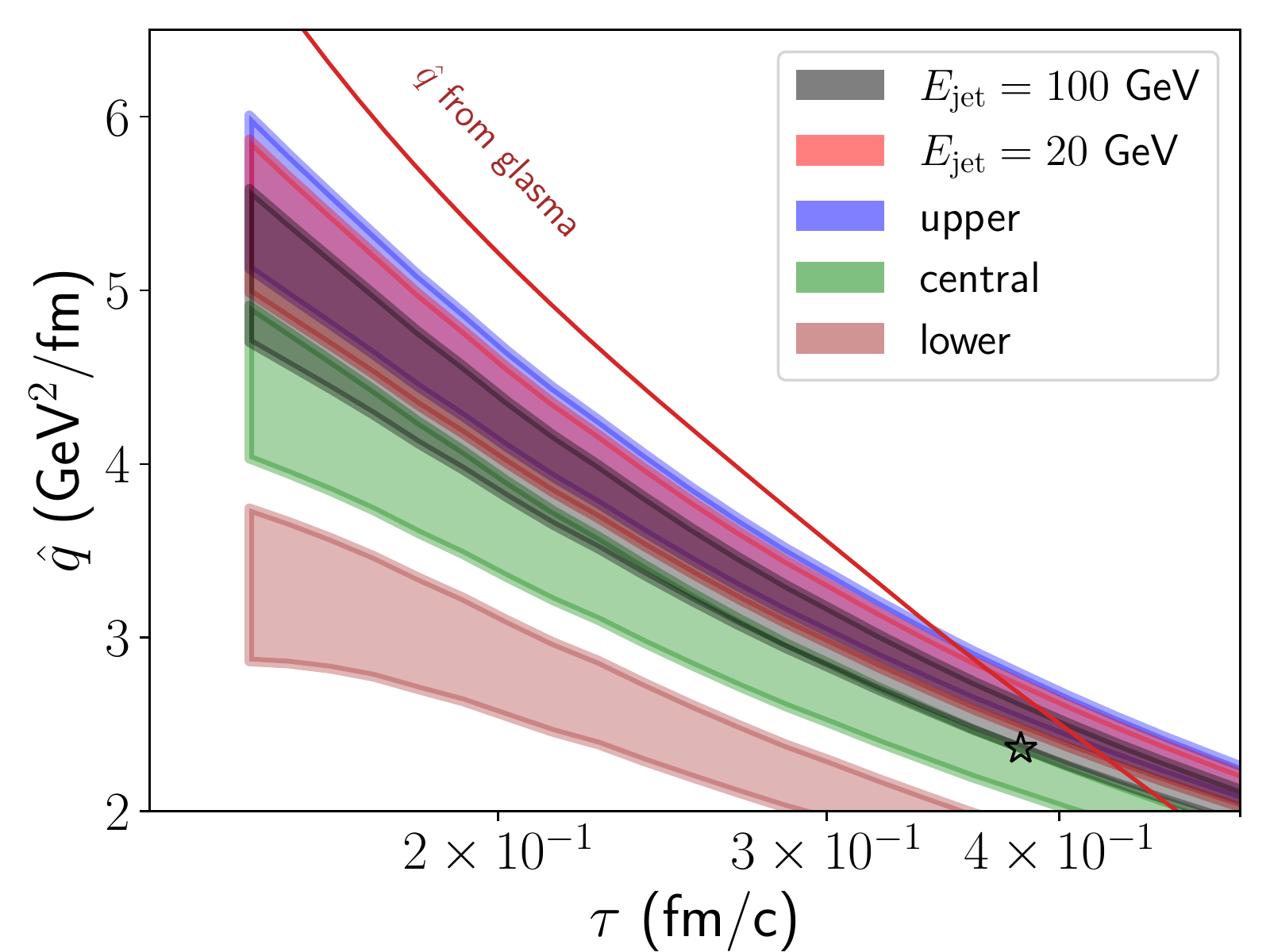}
    \caption{Evolution of $\hat q$ (jet energy $\SI{50}{\giga\electronvolt}$) for different parameterizations according to Ref.~{\cite{Xie:2022ght}}, as explained in the text.}
    \label{fig:different_parametrization_included}
\end{figure}

In the main text, we match the proportionality constants $\zeta$ in {\eqref{eq:cutoff_combined}} with the LBT parametrization from the JETSCAPE result {\cite{JETSCAPE:2021ehl}} for $\qhat$. Here, we investigate possible changes of our description due to a matching with the values from Ref.~{\cite{Xie:2022ght}} for comparison. Their curves are consistent with the LBT model employed in the main text at our matching temperature $\Teps={\SI{295}{\mega\electronvolt}}$. However, to quantify the uncertainty in the extraction, we choose upper ($\qhat = 2.5 T^3$), central ($\qhat = 2 T^3$), and lower ($\qhat = 1.5 T^3$) parts of the error band in Fig.~3 from Ref.~{\cite{Xie:2022ght}}. 
While these values are independent of the jet energy, we choose $\Ejet={\SI{50}{\giga\electronvolt}}$ for Eqs.~{\eqref{eq:cutoff_combined}} in our paper.
We find that the upper and central values are compatible with the JETSCAPE parametrization and also consistent with the glasma values, whereas the lower bound yields slightly smaller values, as visible in \fig{\ref{fig:different_parametrization_included}}, where we have zoomed into the early-time evolution.

\section{Anisotropy of the jet quenching parameter}
\label{app:aniso_qhat}

\begin{figure}
    \centering
    \includegraphics[width=\linewidth]{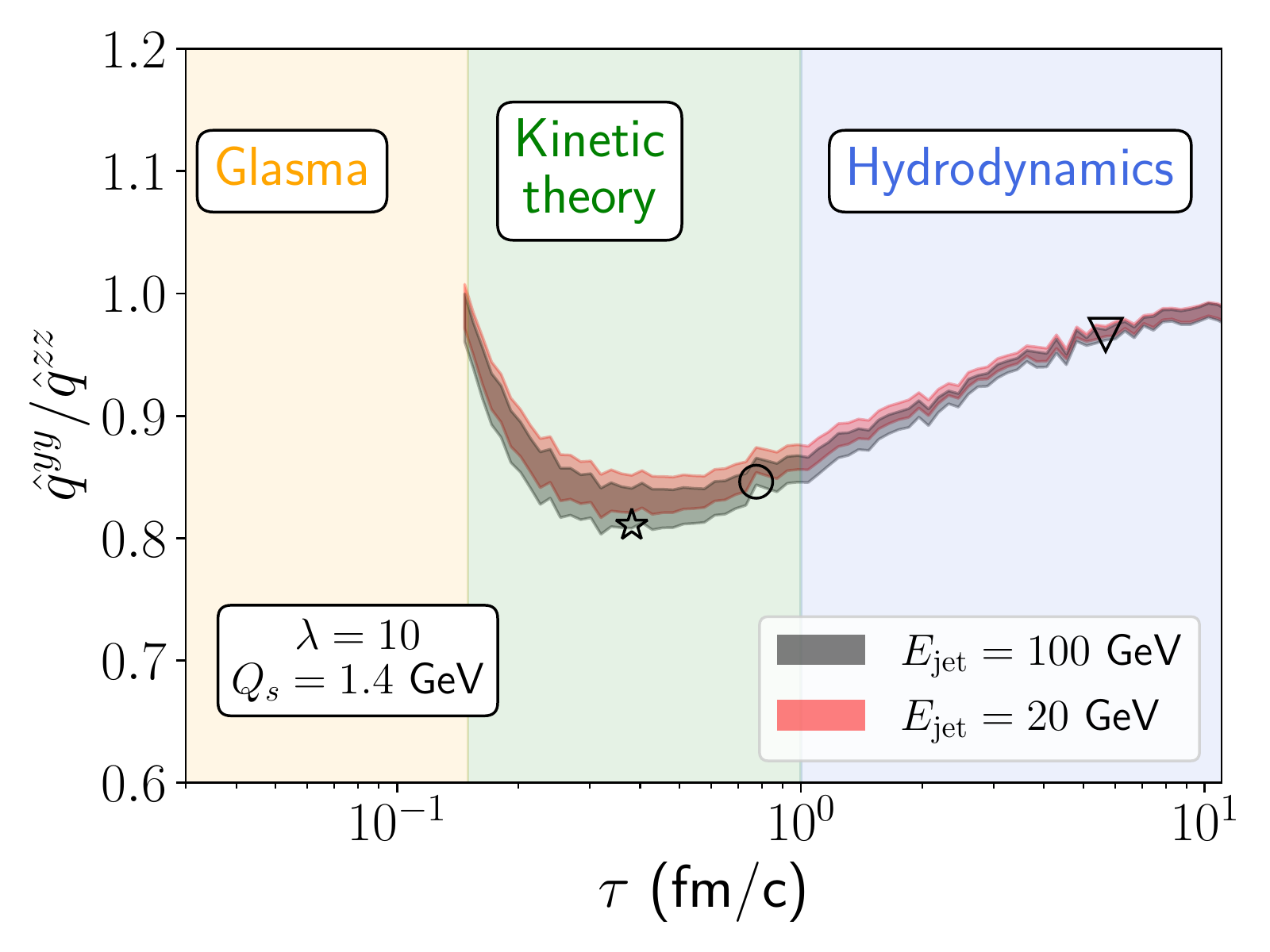}
    \caption{Resolving the anisotropy in \fig\ref{fig:qhat_appetizer_combinedv2} by considering the ratio $\qhat^{yy}/\qhat^{zz}$.}
    \label{fig:anisotropy2}
\end{figure}
In the right panel of \fig\ref{fig:qhat_appetizer_combinedv2}, we plot the absolute value of $\qhat=\qhat^{yy}+\qhat^{zz}$ to make a comparison with the glasma. For phenomenological purposes (see e.g. {\cite{Hauksson:2023tze}}) also the anisotropy is important. We thus show in {\fig{\ref{fig:anisotropy2}}} the ratio $\qhat^{yy}/\qhat^{zz}$ during the hydrodynamization process for each simulation. As in the right panel of \fig{\ref{fig:qhat_appetizer_combinedv2}}, we use jet energies $\Ejet = {\SI{20}{\giga\electronvolt}}$ and ${\SI{100}{\giga\electronvolt}}$, and each band contains simulations with both cutoff models and initial conditions. 
We observe a sizable anisotropy of up to 20\% around $\tau \sim 0.4\, {\si{\femto\meter/\c}}$, before it eventually approaches unity.

\bibliographystyle{elsarticle-num}
\biboptions{sort&compress}
\bibliography{qhatshort}


\end{document}